\documentclass[epj]{svjour}

\usepackage{graphicx}
\usepackage{fancyhdr}
\usepackage{epstopdf}
\def\makeheadbox{{
 }}
\def\mytitle{Indirect Searches for Dark Matter with AMS-02}
\def\myauthors{Pierre Brun}
\def\mytype{}
\def\mysession{}

\begin{document}

\title{Indirect Searches for Dark Matter with AMS-02}

\author{Pierre Brun
\thanks{\emph{Email:}pierre.brun@cea.fr}
}

\institute{DSM/IRFU, Service de Physique des Particules, CEA/Saclay, F-91191
Gif-sur-Yvette, France}
\date{}
\abstract{AMS-02 is a multi-purpose spectrometer with superconducting magnet,
and is designed for 3 years of data taking aboard the International Space
Station. Its high performance regarding particle identification and energy
measurement will allow performing indirect searches for dark matter (DM) in
different channels simultaneously: gamma rays, positrons and antiprotons.
\hbox{AMS-02} sensitivity to those signals are presented and --provided the
positron excess is due to DM signal-- it is shown that it allows to probe new
physics models in detail. Its high sensitivity could even be a unique
opportunity to reach the Majorana nature of the DM particle through final state
polarization effects.
}
\maketitle
\section{Introduction}
\label{intro}

Combinations of different cosmological probes indicate the presence in the
Universe of non baryonic dark matter (DM), at the level of 84\% of the total
matter density. In an independent way, several models beyond the Standard Model
of particle physics (with supersymmetry or extra spacetime dimensions) predict
the existence of neutral, weakly-interacting massive particles~\cite{PhysRep}.
Those could fill the Universe up to the observed matter density provided their
co-moving density is regulated by primordial self-annihilations.

Nowadays, these annihilations could take place in the center of massive
structures containing DM, like {\it e.g.} the center of our Galaxy or in
Galactic clumps of DM. If efficient enough, annihilations of DM particles can
become observable sources of primary cosmic rays. Due to a reduced standard
background, anti-particles ($\rm e^+$, $\rm \bar{p}$, $\rm \bar{D}$) are
amongst the most promising channels, together with $\gamma$ rays. In this
article, we focus mainly on the charged channels.

\section{The AMS-02 space spectrometer}
\label{sec:2}

AMS-02 is a particle physics detector designed for 3 years of data taking in
the GeV-TeV region~\cite{ams02}. The core of the instrument is a Silicon
tracker, surrounded by a superconducting magnet and veto counters. The spatial
resolution is $10\,\rm{\mu m}$ on 8 planes, and the value of the magnetic field
is $B\cdot L^2=0.86\,{\rm T\cdot m^2}$. Two scintillator planes allow to
measure the time of flight with a 120 ps precision and an imaging \v{C}erenkov
counter determines the charge and velocity within $\sim0.1\%$. Proton rejection
is performed with a transition radiation detector and an electromagnetic
calorimeter, each of them providing a $\sim 10^{2-3}$ level proton rejection.
The calorimeter also measures the energy of electromagnetic particles with a
few percent precision.

For what concerns DM channels ($\rm e^+$, $\rm \bar{p}$, $\rm \bar{D}$ and
$\gamma$), the main challenge is the rejection of the important proton
background. Tab.~\ref{tab:1} summarizes the mean expected flux ratio for these
channels with respect to proton background. Specific event selections, using
all sub-detectors, allow the observation of those particles with a high level
of background rejection. As a result of optimized event selections in Monte Carlo simulations, Tab.~\ref{tab:1} shows the obtained acceptance ratios -- $\mathcal{A}_{channel}/\mathcal{A}_{proton}$, that quantifies the capability of the detector to disentangle between the two types of particles. Also shown are the energy ranges for which background contamination is less than 1\%~\cite{jon,caraffini,loic}. Although this paper does not present $\gamma$ channel in AMS-02, we emphasize the fact that its sensitivity to $\gamma$ signal is very high up to a few TeV~\cite{loic,gamma}, in particular with a specific trigger based on the calorimeter signals only~\cite{trigger}.

\begin{table}
\caption{Background rejections and sensitivity ranges.} \label{tab:1}
\begin{tabular}{llll}
\hline\noalign{\smallskip}
 & $\rm e^+$ & $\rm \bar{p}$ & $\rm \gamma$\\
\noalign{\smallskip}\hline\noalign{\smallskip}
$\rm \langle\Phi{channel}/\Phi{protons}\rangle$ & $10^3$ & $10^4$ & $10^{3-5}$ \\
Mean acceptance ratio & $5.6\times 10^5$ & $10^6$ & $10^{4-7}$\\
Energy range (GeV)& 1-300 & 1-600 & 2-2000\\
\noalign{\smallskip}\hline
\end{tabular}
\end{table}

\section{Derivation of DM indirect signals} \label{sec:2}

Apart from the misidentified particles, the other type of background to face
when searching for dark matter is related to the conventional production of
cosmic rays in the Galaxy. Those are presumably produced either in Supernovae
(primaries, which are also accelerated) or by interaction in the interstellar
medium (secondaries). In the case of $\rm e^+$ and $\rm\bar p$, secondaries are
dominant at the Earth in the energy range of interest (1 GeV to a few TeV). If
occurring, DM annihilations in the halo constitute an exotic production of
primary anti-particles. The flux of DM-induced charged particles at the Earth
$\Phi$ is driven by a transport equation of the form

\begin{equation}
\partial_{z}(V_c \Phi) - K\Delta\Phi + \partial_{E}\left\{ b(E) -
K_{\rm EE}(E)\ \partial_{E} \Phi \right \} = \mathcal{Q}(\vec{x},E).\label{eq:1}
\end{equation}

Here, $V_c$, $K$, $b(E)$ and $K_{\rm EE}$ are parameters of the propagation
model and respectively stand for the convective wind, the diffusion
coefficient, the energy loss rate and the diffusive reacceleration term.
$\mathcal{Q}$ is the source term and is discussed below. This equation is
solved with cylindrical boundary conditions that correspond to the expected
shape of the diffusion zone. Depending on the considered species, some terms
can be neglected (like {\it e.g.} $b(E)$ for antiprotons) and different methods
can be applied for the resolution\cite{david,julien}.

The source term, that contains the new physics and the modelling of the DM
halo, can be expressed as

\begin{equation}
\mathcal{Q}(\vec{x},E) \;\;= \;\;\frac{1}{4\pi}\;\;\frac{\langle \sigma
v\rangle}{m_{\chi}^2}\;\; \frac{dN}{dE}(E)\;\;
\frac{\rho_{\chi}^2(\vec{x})}{\delta}\;\;,\label{eq:2}
\end{equation}
where $\langle \sigma v\rangle$ is the thermally-averaged annihilation
cross-section, $m_{\chi}$ is the DM particle mass, $dN/dE$ is the energy
spectrum produced in the annihilation (which is discussed below) and
$\rho_{\chi}$ is the DM density. $\delta$ is a statistical factor, equal to 2
for identical particles and 4 for a mixing of two populations ({\it e.g.} DM
and anti-DM particles). Additional negative source terms can account for the
disappearance of the particles by interaction in the Galactic disc.

Each pair annihilation of DM particles leads to the production of a pair of
standard particles. The types and flavors of the produced particles are
associated to a probability which is determined through the new physics model.
As it is shown in the following, a possible determination of the final state
mean $dN/dE$ spectrum can lead to the identification of the underlying theory.

It is important to notice that the presence of DM substructures could
significantly enhance the annihilation rate. The density of those can either
follow the overall DM density (clumps) or not, like {\it e.g.} in the case of
mini-spikes around intermediate-mass black holes (IMBHs). Indeed, the rate
being $\propto\rho^2$, the higher the variance of the DM distribution, the
higher the exotic signal, as a clumpy distribution would lead to a boost $\propto \langle \rho^2 \rangle/ \langle \rho \rangle ^2$ with respect to a smooth one. Although enhancement factors can be explicitly
computed within given substructure scenarios~\cite{julien,imbhs}, we will here
adjust the value of the enhancement (in fact, the distance at the nearest
substructure) in order to reproduce the so-called positron
excess\cite{heat1,heat2,ams01}. In the following, the type of substructure will
not be specified, but maybe IMBHs type should be preferred since the idea of
sizeable clumpiness enhancement have recently been doubted~\cite{julien2}.

\section{Prospects for AMS-02} \label{sec:3}

\subsection{Supersymmetry and Extra-dimensions signals}

In this section, we consider two models for the DM particle, within
supersymmetry (SUSY) and Kaluza-Klein (KK) scenarios with extra-dimensions. In
the two cases, specific models and parameters are chosen to exhibit typical
features of each scenario. In both cases, candidates of 150 GeV mass are
considered with sets of parameters that match cosmological constraints on the
relic density. The computations of the exotic fluxes --including
cross-sections, dark halo modelling and propagation-- are performed with a new
version of the~\verb"micrOMEGAs" package~\cite{micro1,micro2}.

In the minimal supersymmetric extension of the Standard Model (MSSM), the
typical candidate is the lightest neutralino (LSP). The SUSY breaking scenario
we consider here is a mSUGRA type with unification of the scalar masses to
$m_0=113$ GeV and of the fermions to $m_{1/2}=$ 375 GeV at an energy of
$\mathcal{O}(10^{19})$ GeV. We also fix $\rm{tg}\beta=50$, $\mu>0$ and a
vanishing trilinear coupling. The mass spectrum is computed with the
\verb"SuSpect" package~\cite{suspect}, and leads to a 150 GeV neutralino. The
$bino$ component being dominant, the annihilation goes through a mixing of
$b\bar{b}$ and $\tau^+\tau^-$ pairs in a $60/40$ proportion. The total
annihilation cross section, as computed with \verb"micrOMEGAs", is $\langle
\sigma v \rangle=7\times 10^{-28}\;\rm cm^2\cdot s^{-1}$.

The KK scenario used here considers warped extra-dimensions with a SO(10)
unification group~\cite{agashe}. The DM particle in this scenario is referred
to as the LZP (in reference to the name of the discrete symmetry that
guarantees proton stability). In this framework, the annihilation occurs mainly
{\it via} the exchange of $Z^0$ or $Z'$ bosons in the s-channel. Considering KK
bosons mass scale of 3 TeV, a proper thermal relic density can be obtained for
a DM particle mass of 150 GeV~\cite{agashe}, with a total annihilation cross
section $\langle \sigma v \rangle=2\times 10^{-26}\;\rm cm^2\cdot s^{-1}$. In
this framework, 74\% of the annihilation produce quark pairs, and the direct
branching fraction into charged leptons is 3\% each. This represents a high
proportion of direct $\rm e^+/e^-$ pairs. The positrons which are produced this
way are particularly interesting since they are mono-energetic.

Let us focus now on the specific features of each type of model. For what
concerns antiprotons, they are exclusively produced in the quark hadronization
processes ($\tau$ channels do not produce $\rm \bar{p}$). The quark yield being
more important for the LZP, a little more $\rm \bar{p}$ are expected in that
case. However, the most important differences between the two spectrum
predictions appear for positrons. Indeed, the relatively large production of
$\rm e^+$ line in the LZP case lead to a spectrum that exhibit a very sharp end
at 150 GeV, which does not appear in SUSY. The scope of this paper is to show
that such spectra can easily be identified with 3 years of data taking with
AMS-02.
\begin{figure*}
\centering
\includegraphics[width=6.6cm]{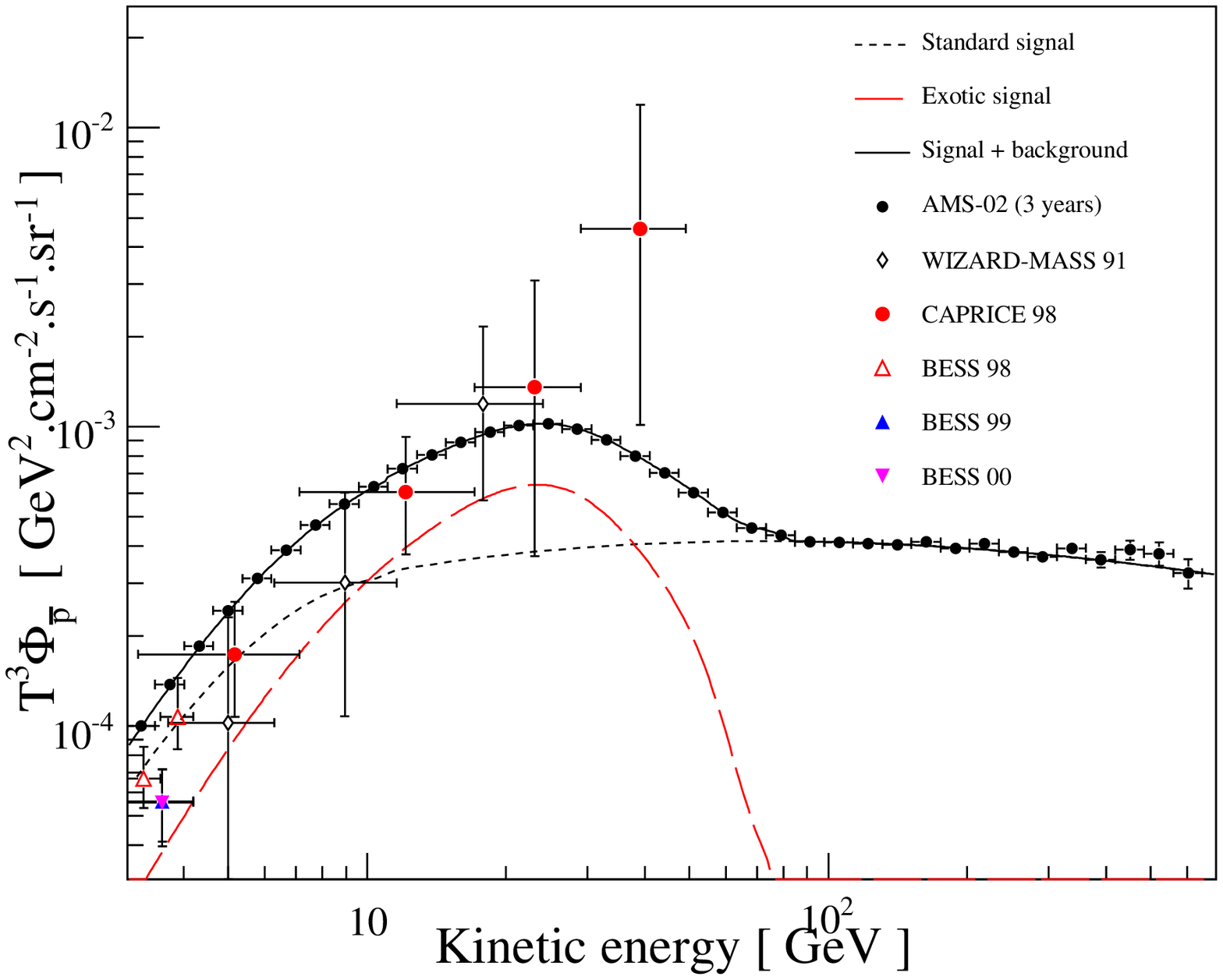}
\includegraphics[width=6.6cm]{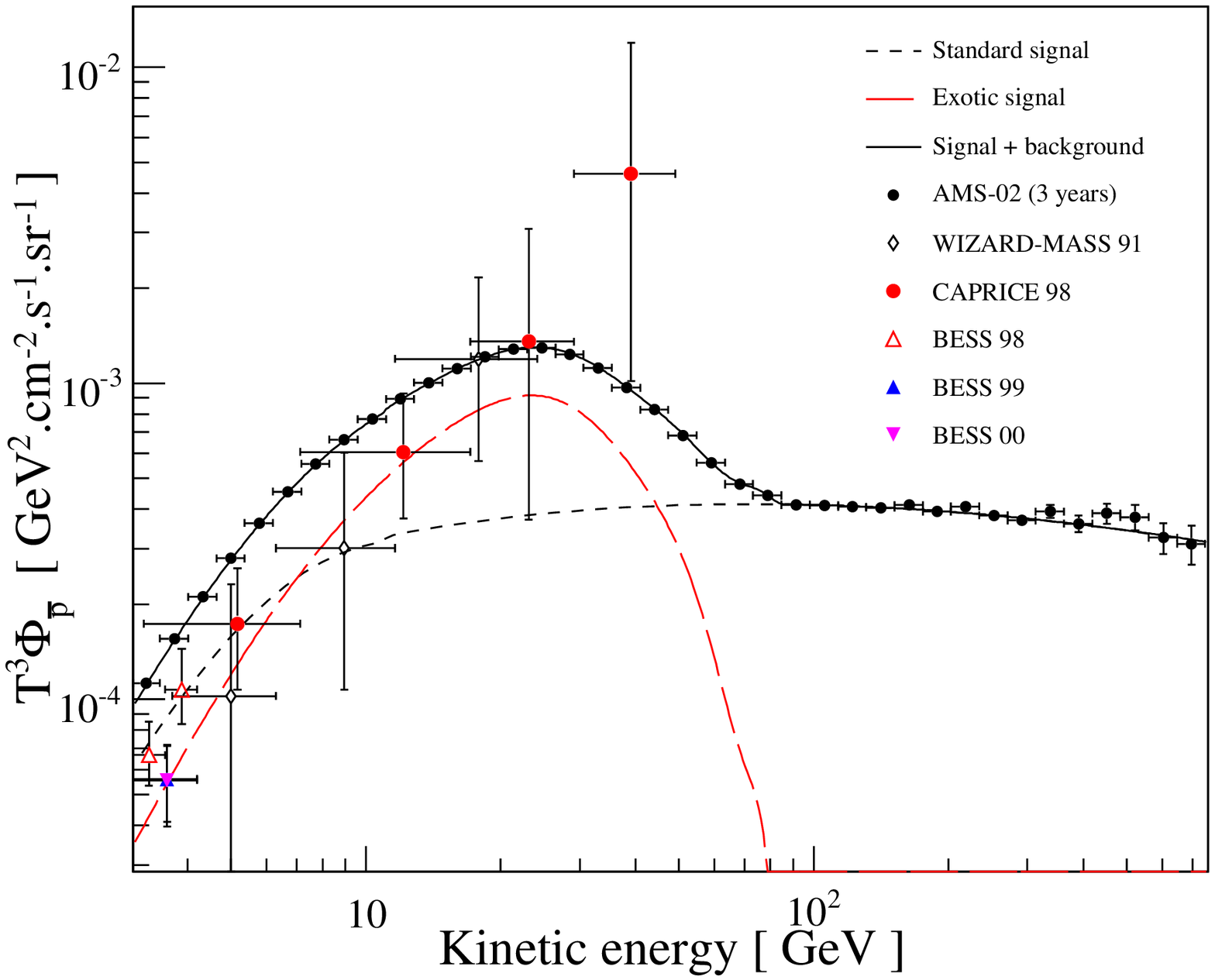}
\includegraphics[width=6.6cm]{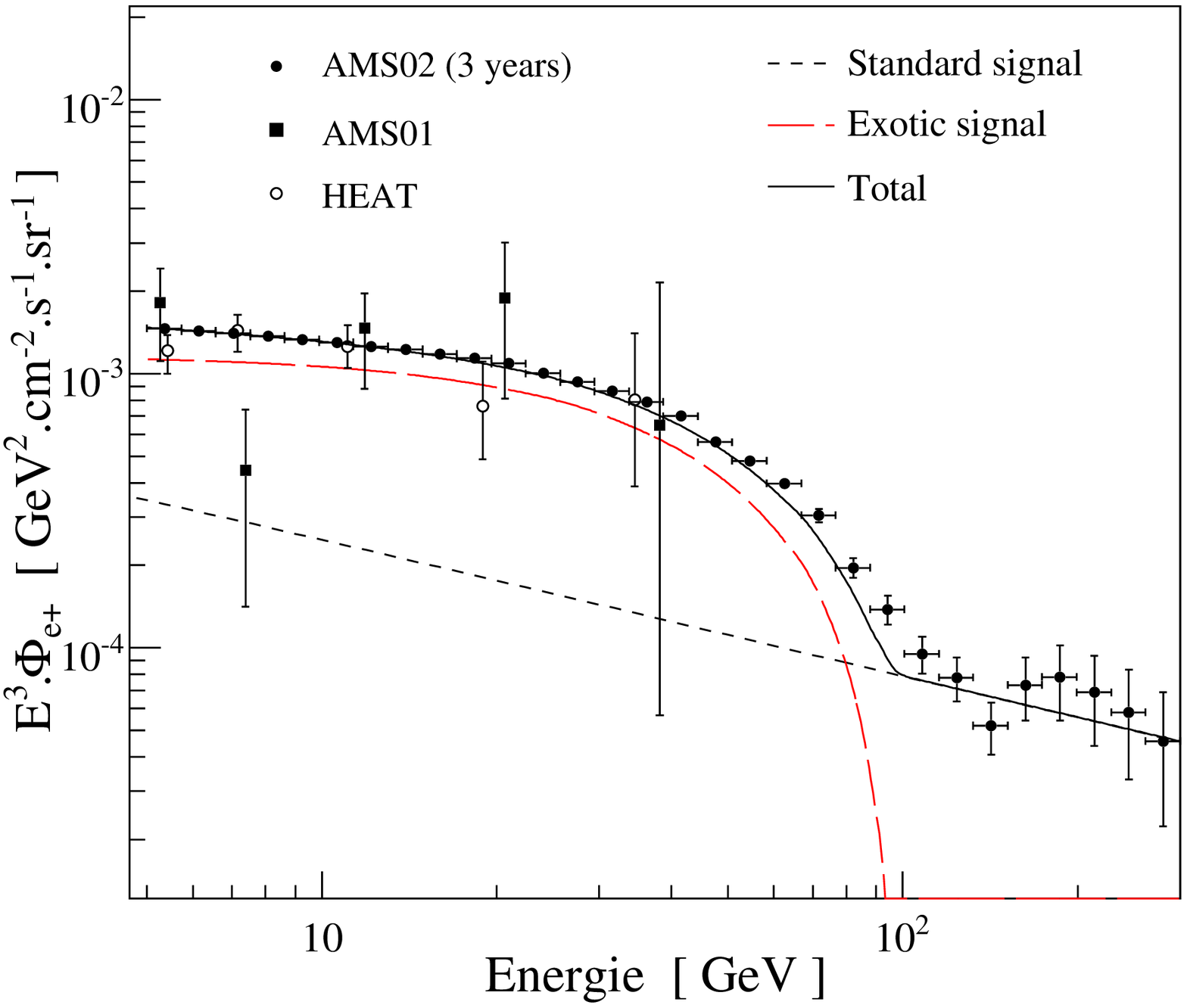}
\includegraphics[width=6.6cm]{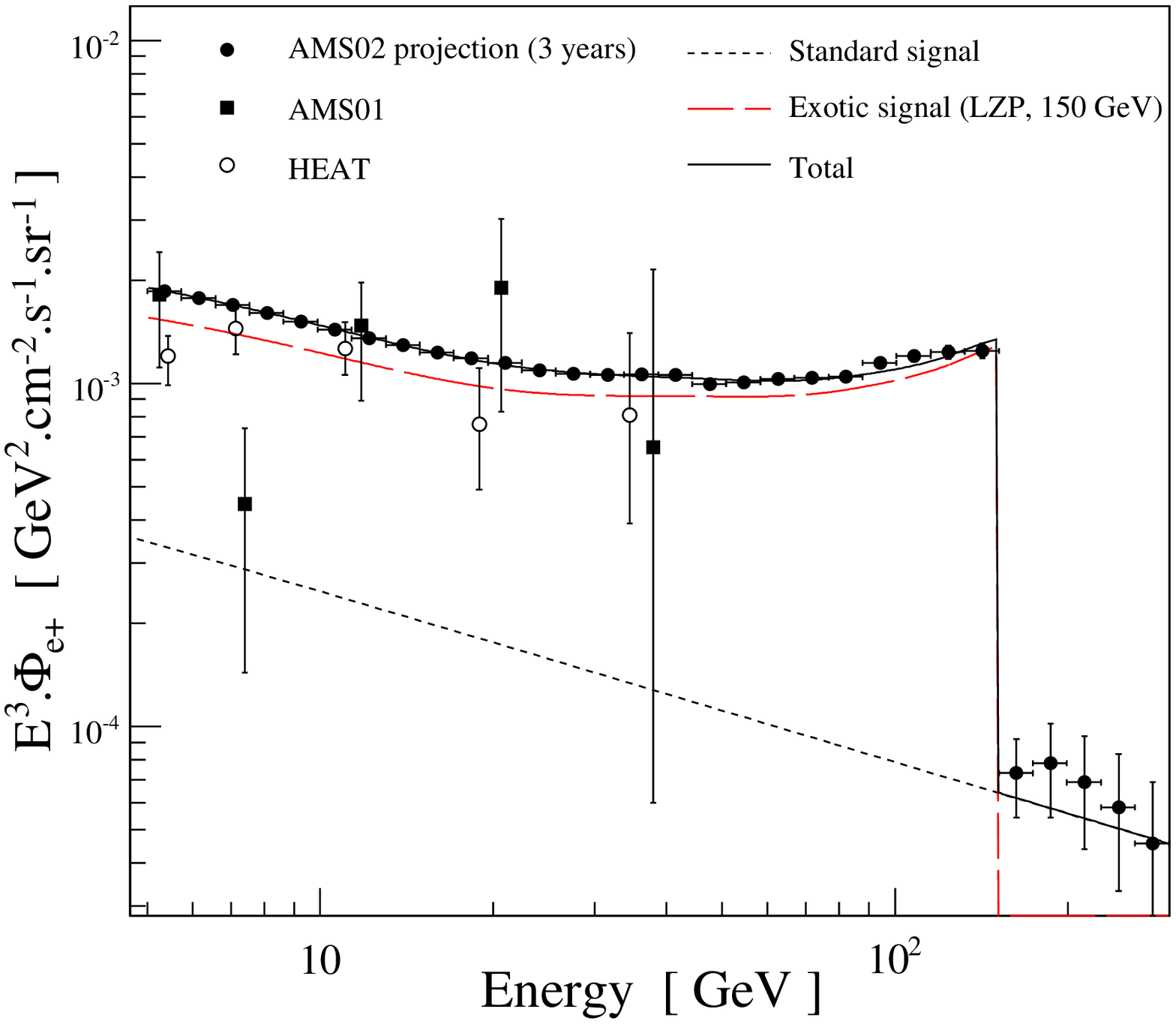}
\caption{\label{fig:1}Expected measurements of the $\rm e^+$  and $\rm \bar{p}$ fluxes for SUSY DM
(a $bino$, left) and a LZP (right).}
\end{figure*}
To do so, we use a comprehensive simulation of the AMS-02 apparatus, and
include all systematics related to misidentification of the different
backgrounds. We include also statistical uncertainties accounting for 3 years
of data taking. In the considered energy range, the theoretical uncertainties
on the flux predictions decreadses as the energy increases. In fact, the higher
the energy, the more purely diffusive the transport regime. As a consequence,
high energy fluxes are more under control since they depend on less parameters.
For that reason, fluxes are only considered above 3 GeV here.

Fig.~\ref{fig:1} displays the different spectra as they would be measured by
AMS-02, together with current measurements. The upper panels present $\rm
\bar{p}$ measurement and the lower panels show the $\rm e^+$ signals. In each
case, the SUSY candidate is considered on the left and the KK one is on the
right. Standard fluxes are determined from~\cite{salati} for $\rm \bar{p}$\footnote{In the case presented here, the standard signal is consistent with BESS results and AMS-02 points are assumed to be slightly higher.} and
from~\cite{baltz} for $\rm e^+$. One can see that, apart from the normalization
of the signals, the $\rm \bar{p}$ signals do not exhibit significant
differences (this degeneracy in the $\rm \bar{p}$ signal has been pointed at in ~\cite{salati}). On the contrary, as mentioned above, the $\rm e^+$ line production
lead to significant dissimilarities in the positron spectra. It appears that
the high sensitivity of AMS-02 to this channel allows to clearly distinguish
the two models. In particular, the sharp cutoff in the KK-induced $\rm e^+$
spectrum can be detected with a high significance.

While we do not show the results for $\gamma$ rays expected measurements here,
it is important to notice also that --unlike the KK case-- the SUSY case leads
to an observable 150 GeV $\gamma$ line. This result is obtained thank to the
high performance of the calorimeter.

Notice finally that, even for those relatively light neutralinos, fluxes
measurements on a large energy range are essential. In fact, only a precise
measurement of the spectra \emph{above} the neutralino mass can lead to an
identification of the expected spectral features. Furthermore, it allows a
direct measurement of the standard secondary background and helps in fitting
shape of the signal. From that point of view also, AMS-02 great advantage is to
offer large energy range measurements.

\subsection{Polarization of final state gauge bosons}

In this section, we consider the specific case of an annihilation into pairs of gauge
bosons ($\chi\chi\rightarrow W^+W^-$). This occurs in the MSSM when the
LSP has a large Higgsino component (like, {\it e.g.} in anomaly-mediated SUSY
breaking scenarios), or within universal extra-dimensions frameworks.

\begin{figure*}
\centering
\includegraphics[width=7.3cm]{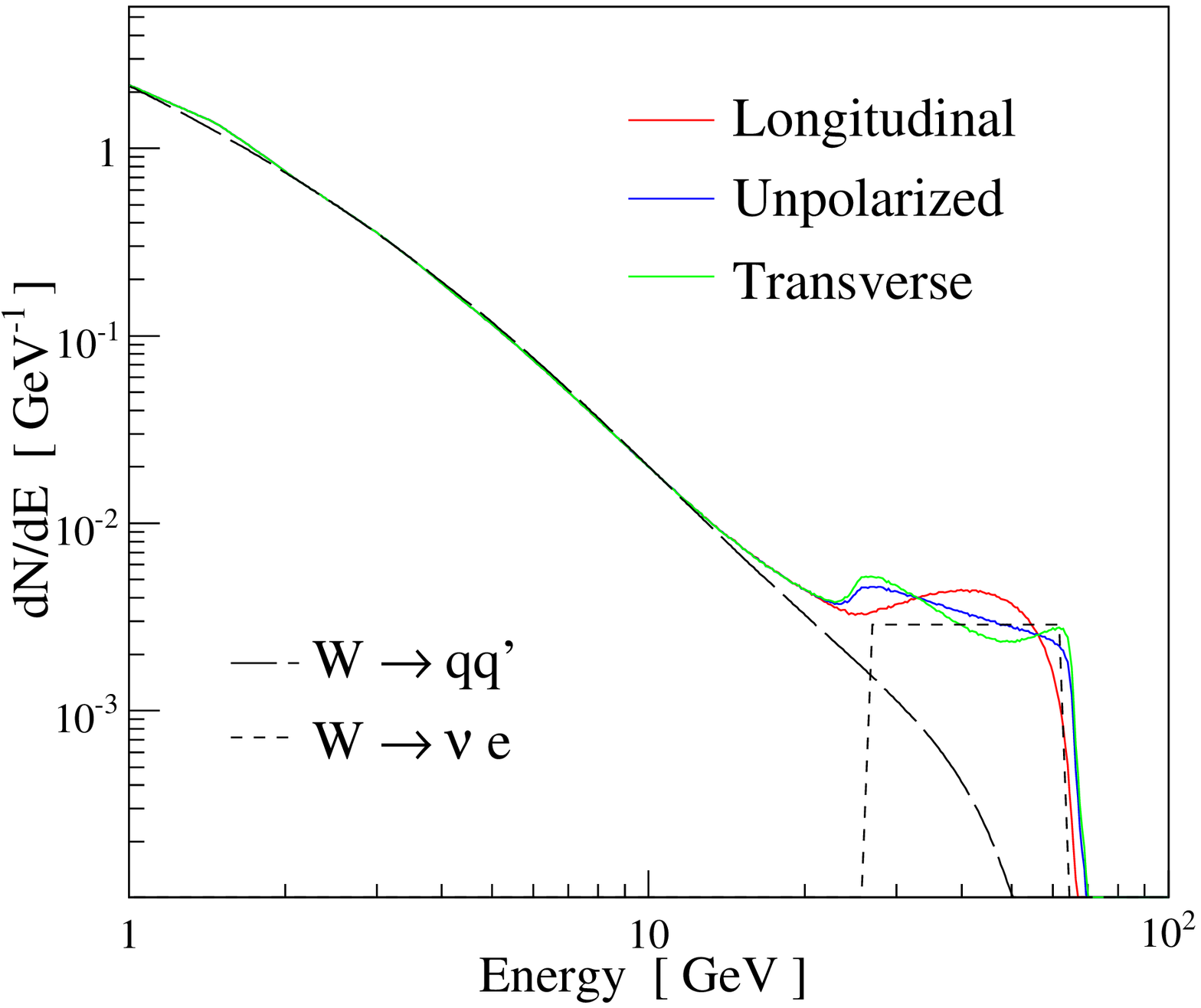}
\includegraphics[width=7.5cm]{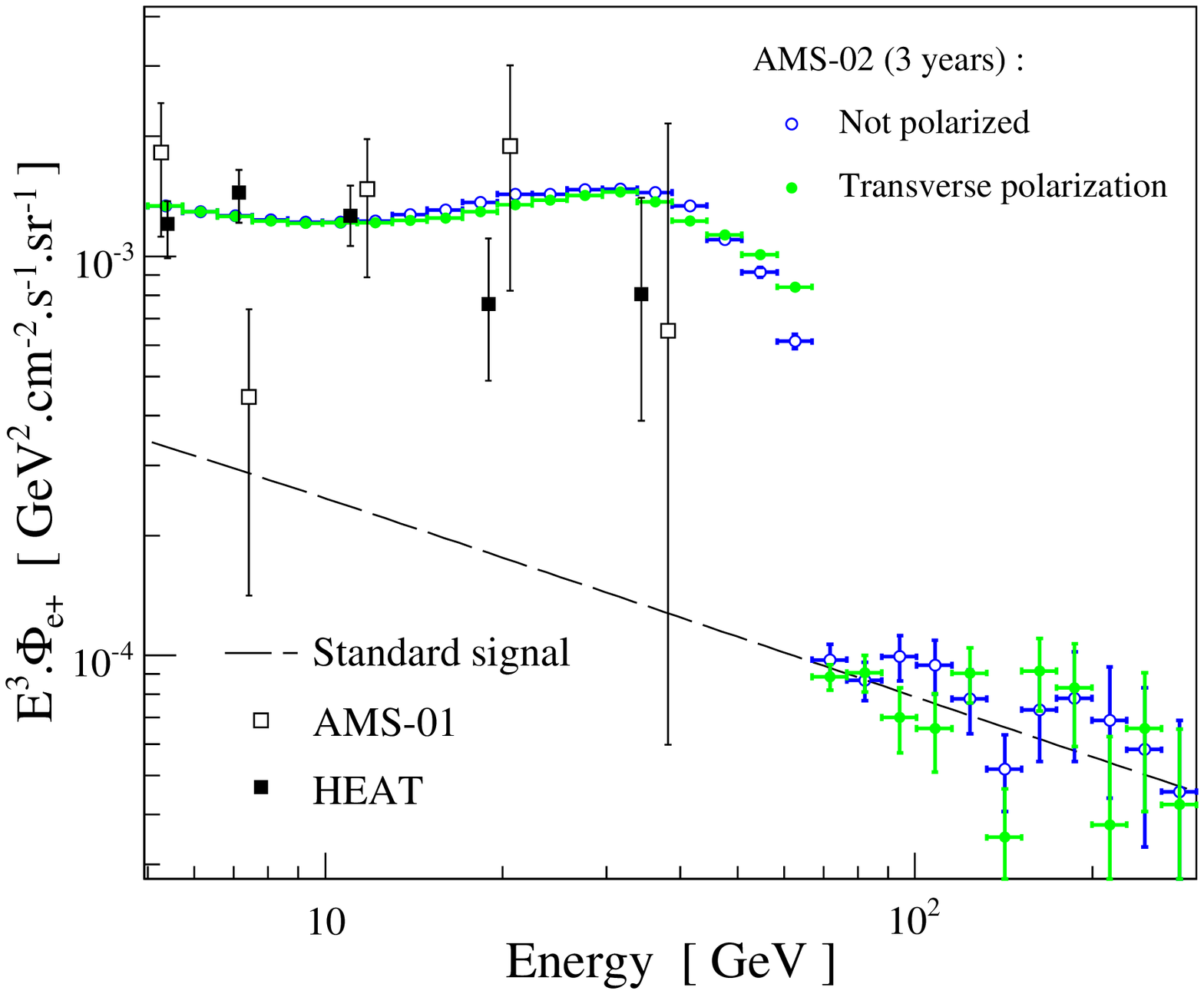}
\caption{\label{fig:2} $\rm e^+$ spectra for different polarization, before propagation (left) and as they would be measured by AMS-02.}
\end{figure*}

We point out here that the energy spectrum of the outgoing $\rm e^+$ (and
$\nu$) depends on the polarizations of the produced gauge bosons. Indeed,
Majorana particles like SUSY neutralinos lead to transversally polarized gauge
bosons, whereas Dirac particles such as KK candidates produce unpolarized
bosons. $\rm e^+$ and $\nu$ spectra should be different in each case, as we
shall see in the following. This effect has never been pointed at before, a
more detailed study will appear in~\cite{wpol}.

The left panel of Fig.~\ref{fig:2} displays the $dN/dE$ functions for a generic
100 GeV DM particle annihilating into $W$ pairs with different polarizations.
The specific shapes of each curve are related to the angular distribution of
the leptons in the $W\rightarrow \ell \nu$ processes, depending on the $W$
polarization. After propagation of these spectra, the differences between the
two cases tend to vanish but we show that a sizeable difference stands when the
fluxes are considered at the Earth.

In the same way as in previous section, the prediction for AMS-02 observations
in 3 years of data taking are performed, assuming the DM signal is the origin
of the positron excess. The right panel of Fig.~\ref{fig:2} shows the expected
AMS-02 measurements in each case. It appears that --under the assumption of a $W$
dominated annihilation (which could be inferred, {\it e.g.}, from the LHC
data)-- a few bins show significant differences, in particular around 20 GeV and in the
last bins of the excess. A precise analysis of the spectral shape of the excess could allow
to distinguish between the two cases through a fit of the data points with hypotheses on the signal shape. Notice that the determination of the Majorana nature of the DM particle would be impossible with any current experiment, neither at the LHC. Only the future linear collider could reach that kind of information.

\section{Conclusions}

AMS-02 is a large acceptance space spectrometer that will perform precise
measurements of cosmic radiation up to the TeV region, with a high level of
background rejection. The prospect concerning indirect searches for dark matter
are very promising in the antiprotons, positrons and gamma rays channels. In
particular, it is shown that not only an excess can be observed, but specific
new physics models can be unambiguously identified. The precise measurement of
the high energy part (above the DM particle mass) insures to simultaneously
have a determination of the standard background. In addition, the accurate
determination of the nuclei fluxes will lead to important constraints on the
propagation models and significantly reduce the theoretical uncertainties, thus
allowing to reach even higher sensitivities to DM models.

On top of that, the high sensitivity to positrons and the precision of the
energy measurement allow to probe very fine details of the new physics
underlying the dark matter problem. For example, if the positron excess is due
to particles annihilating in gauge boson pairs, AMS-02 as enough sensitivity to give clues
on the Majorana structure of the dark matter particle.

\end{document}